\documentclass[english, 12pt]{article}

\usepackage[T1]{fontenc}
\usepackage{babel}
\usepackage{amsmath}
\usepackage{graphics}

\newcommand\lsim{\mathrel{\rlap{\lower4pt\hbox{\hskip1pt$\sim$}}
    \raise1pt\hbox{$<$}}}
\newcommand\gsim{\mathrel{\rlap{\lower4pt\hbox{\hskip1pt$\sim$}}
    \raise1pt\hbox{$>$}}}

\begin{document}

\title{Vorton Existence and Stability}

\author{Y. Lemperiere\footnote{E-mail:
    Y.F.J.Lemperiere@damtp.cam.ac.uk} \, and E.P.S. Shellard\footnote{E-mail:E.P.S.shellard@damtp.cam.ac.uk}}

\date{\emph{DAMTP, Wilberforce street, Cambridge CB3 0EW, United-Kingdom}}

\maketitle

\begin{abstract}

We present the first concrete evidence for the classical stability of
vortons, circular cosmic string loops stabilized by the angular momentum 
of the charge and current trapped on the string.   We  begin by summarizing 
what is known about vorton solutions and, in particular,
their analytic stability 
with respect to a range of radial and nonradial perturbations.  
We then discuss numerical results of 
vorton simulations in a full 3D field theory, that is, Witten's original 
bosonic superconducting string model with a modified potential term.
For specific parameter values, these simulations demonstrate the 
long-term stability of sufficiently large vorton solutions 
created with an elliptical initial ansatz.

\end{abstract}

PACS: 98.80.-k;74.60.Jg

\section{Introduction}

Topological defects are known to play an important role in many physical 
contexts and they may also impact cosmology, the relativistic setting
for this present study (for a review, see ref.~\cite{VS:CS}).  Amongst the
possible cosmic defects, vortex-strings have a prominent position because they
naturally evolve into a scale-invariant configuration, therefore
avoiding analogues of the monopole problem of cosmological domination. 
These strings might be responsible for
a variety of astrophysical phenomena, such as cosmic rays,
gravitational wave radiation, or gravitational lensing. The richness of their
phenomenology comes in part from the possibility of additional internal
structure, making them superconducting ref.~\cite{WI:SCS}.

These superconducting strings have been widely studied, and this article
follows a companion paper \cite{LS:VO}, which investigated in detail
the model studied
here.  Previously, we considered analytic criteria for current and charge
stability on these strings, confirming this with numerical
simulations.  This work 
allowed us to predict the equilibrium state of a cosmic string loop
stabilized by 
the angular momentum of this charge and current, while also setting
new limits on 
the available range of parameters.  Let's mention that the potential
existence of vorton 
solutions has been postulated recently in other important physical
contexts such as 
QCD \cite{BMZ:QCD} and high-$T_{\rm c}$ superconductivity \cite{BZ:SO5}.  
We note that if we are able, here, to establish vorton stability in vacuum 
(i.e. resisting collapse due to the powerful tension of a relativistic string),
then these results bode well for their as yet untested stability in less 
extreme physical situations. We emphasise 
that here we are focussing on classical vorton stability, while for a
discussion of quantum stability the reader 
is referred to \cite{BPOV:SCS} and references therein. 

The structure of this letter is as follows:
After summarizing our previous results for vorton equilibrium states, 
we will examine additional
relevant perturbations which we have observed as a result of our work.
Finally, we will give an account of our numerical results on vortons and
their long-term stability given elliptically perturbed initial conditions.

\section{The framework}

Our study is based on the model first proposed in ref.~\cite{WI:SCS}~:
\begin{eqnarray}
\mathcal{L} &= (\partial_{\mu}\phi)(\partial^{\mu}\phi)^+ +
(\partial_{\mu}\sigma)(\partial^{\mu}\sigma)^+ -
\frac{\lambda_\phi}{4}(|\phi|^2-\eta_\phi^2)^2 \nonumber\\
&~~~~~~~~~~- \frac{\lambda_\sigma}{4}(|\sigma|^2-\eta_\sigma^2)^2 -
\beta|\phi|^2|\sigma|^2 \, ,
\label{lagrangian}
\end{eqnarray}
where the two complex scalar fields $\phi$ and $\sigma$ are minimally 
coupled, each being invariant under $U(1)$ transformations.

If the constants of the theory (the couplings and the vacuum
expectation values) are chosen carefully, one can break the
$\phi$-symmetry, leading to $|\phi| = \eta_{\phi}$, while keeping
$\sigma = 0$ in the vacuum, because of the non-vanishing interaction
term. Along a $\phi$ cosmic string, the interaction vanishes, and
$\sigma$ can form a condensate. (For a detailed acount on this model,
see ref.~\cite{LS:VO})
We have been able to give necessary conditions on the parameters of the
lagrangian for the string to be superconducting:
\begin{eqnarray}
\lambda_{\phi} < \frac{\lambda_{\sigma}^2}{16\beta}, \quad \quad & \beta <
\frac{\lambda_{\sigma}}{4}, \quad \quad & \eta_{\sigma}^2 <
\frac{\eta_{\phi}^2}{2}. \quad \quad
\,.
\label{condparam}
\end{eqnarray}

This $|\sigma|$ condensate can also carry charge and currents along
the string (taken to lie on the $z$-axis), as can be
readily seen from the ansatz:
\begin{equation}
\sigma = |\sigma|(r)e^{i(\omega t +kz)}\,,
\label{field}
\end{equation}
which induces a (Noether) charge $Q$ and a current $J$ on the string
worldsheet:
\begin{equation}
Q = \omega \int dz \int dS\, |\sigma|^2 \, , \qquad\qquad J = k \int
dz \int dS\, |\sigma|^2 \, ,
\label{charges}
\end{equation}
as well as a topologically conserved quantity, the winding number:
\begin{equation}
N = \int dz\, \frac{k}{2\pi} \, .
\label{N}
\end{equation}
These features alter significantly the standard string cosmology
scenario, with perhaps the most striking consequence of
superconductivity being the
classical stability of string loops. These loops, which 
were dubbed \emph{vortons} in ref.~\cite{DS:SCS1,DS:SCS2}, cannot decay
because of the angular momentum of their charge carriers.

\begin{figure}
\resizebox{13cm}{6.5cm}{\includegraphics{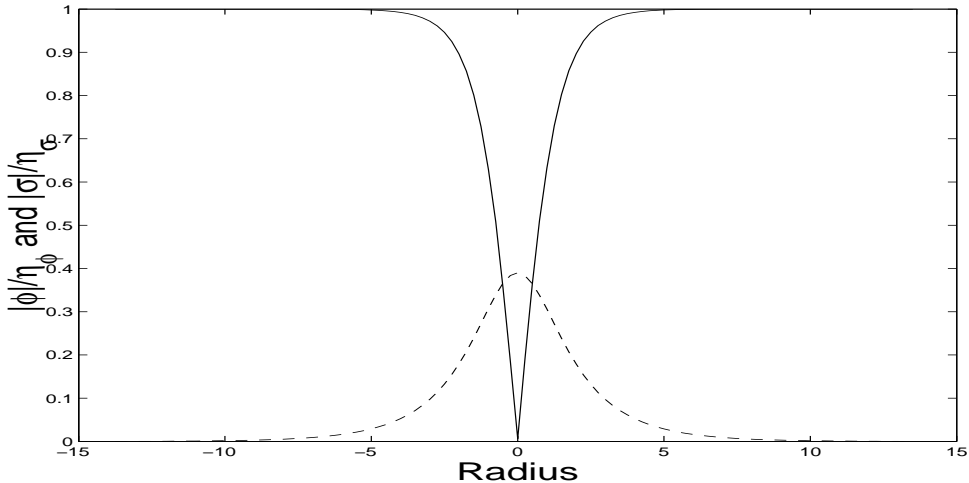}}
\caption{Profiles of $\phi$ and $\sigma$, plotted in units of
$\eta_{\phi}$ and $\eta_{\sigma}$; the parameters here are
$\lambda_{\phi} = 1.5$, $\eta_{\phi} = 1.0$, $\lambda_{\sigma} = 10.0$,
$\eta_{\sigma} = 0.5$, $\beta = 1.5$}
\label{phi2}
\end{figure}

We present in fig.\ref{phi2} a typical, straight superconducting
string profile. The
parameters used here saturate the bounds given by
eq.(\ref{condparam}), in order to get a condensate
well localised along the string, appropriate for numerical
simulations with boundary restrictions. In general, one can expect a
condensate somewhat broader
than the underlying vortex: $\delta_{\sigma} > \delta_{\phi}$, or
equivalently, $\beta\eta_{\phi}^2 -
\frac{1}{2}\lambda_{\sigma}\eta_{\sigma}^2 = m_{\sigma}^2 < m_{\phi}^2
= \frac{1}{2}\lambda_{\phi}\eta_{\phi}^2$. It is probably worth
emphasizing here the model dependence of vorton studies: the parameter
space is very broad, and we need a specific particle physics model fixing the
couplings to make more precise predictions about their properties.

In our previous paper, we have been able to analyticaly characterize
the equilibrium states proving, in principle, that vortons should
occur for every
initial non-zero value of $Q$ and $N$. If, as usual, we call $\mu$ the string
tension, and we define
\begin{equation}
\Sigma = \int dS \, |\sigma|^2 \,, \quad \quad \Sigma_4 = \int dS \,
|\sigma|^4 \,,
\label{defs}
\end{equation}
we can recall the following equilibrium conditions. In the
\emph{chiral} case $\omega^2
= k^2$, vortons will shrink or expand until:
\begin{equation}
\omega^2 = k^2 = {(\mu -\frac{1}{4}\lambda_{\sigma}\Sigma_{4o})}/({2
\Sigma_o}) \, ,
\label{chiral}
\end{equation}
where the suffix $o$ denotes the chiral value of a quantity (or
equivalently, when $\omega = k = 0$). In the\emph{ electric} ($\omega^2 >
k^2$) or\emph{ magnetic} ($\omega^2 < k^2$) regimes, given that $\Sigma_{QN}
= Q/N$, we showed that the vorton state would minimize the following function
of $\omega^2 - k^2 = u^2$:
\begin{equation}
\mathcal{E} =N \left[ \frac{(\mu  -
\frac{\lambda_{\sigma}}{4}\Sigma_4(u^2))}{\Sigma(u^2)}
\sqrt{\frac{\Sigma_{QN}^2-\Sigma^2(u^2)}{u^2}} + 2
\Sigma_{QN}^2 \sqrt{\frac{u^2}{\Sigma_{QN}^2-\Sigma^2(u^2)}} \right] \,,
\label{nentot}
\end{equation}
which clearly admits such a minimum.  Given an initial state with small 
seed charges and currents, we demonstrated that the smaller final equilibrium 
state will generically be located away from the chiral state (which is
{\it not} an attractor -- see ref.~\cite{LS:VO} for details).

\section{Stability analysis}

We employed numerical methods to investigate the behaviour of these equilibrium
vorton states, which
enabled us to identify and analytically characterise various instabilities 
corresponding to
different regimes of the loop. 
We have already proved in \cite{LS:VO} that a straight superconducting string
cannot evolve too deeply into the magnetic regime ($\omega^2 < k^2$),
because of nonlinear effects that force the condensate to become
pinched locally and to unravel, thus losing winding number. 

Our analytical analysis, and a suitable ansatz for $\Sigma$ and
$\Sigma_4$, gave a precise criterion for the threshold value $k_{inst}$ of this
instability. Let's define $\alpha =
\frac{\lambda_{\sigma}}{4}\Sigma_{4o}/\Sigma_o$, and $k_c$, the
maximum $k$ value associated with a non-vanishing condensate. Then, 
\begin{equation}
\frac{1}{k_{inst}^2} = \frac{1}{\alpha} + \frac{1}{k_c^2}
\end{equation}

We have also studied the chiral and electric cases, and
we were unable to find an instability of this kind, suggesting that these
regimes are stable against a `pinching' perturbation.
There is, however, another potential instability particularly relevant in the 
electric regime.

In the model presented
in eq.~(\ref{lagrangian}), the radial loop equilibrium discussed above
in \ref{nentot} is only valid if one assumes that the charge and the
current remain localised on the string's worldsheet.  When 
we consider small loops, especially those tractable
numerically, this assumption is not necessarily valid.
The potential cost of the vortex/condensate separation can be
evaluated using the usual profiles for the vortex, and we obtain the 
following stability
criterion (see ref.~\cite{LS:DCS}):
\begin{equation}
\frac{R}{\delta_{\sigma}} \gsim
\frac{\lambda_{\phi}}{2\beta}\frac{\mu_{\phi}}{\sigma_o^2} \, ,
\label{condf}
\end{equation}
where $R$ is the radius of the vortex loop.
Generically, the right hand side of (\ref{condf}) is quite large, and
so stable loops require $R >>
\delta_{\sigma}$. This is a problem for real
vortons in the electric regime. The width of the condensate
diverges as they go deeper into the electric regime, and so
these loops can prove quite hard to stabilise. Numerically, memory limitations
are such that we cannot
achieve more than $R \sim 20 \delta_{\sigma}$, and thus we are faced
with this potential splitting instability even in the chiral regime,
since this is just the order of magnitude given by (\ref{condf}), as
we have demonstrated in our simulations.

\begin{figure}
\resizebox{13cm}{6.5cm}{\includegraphics{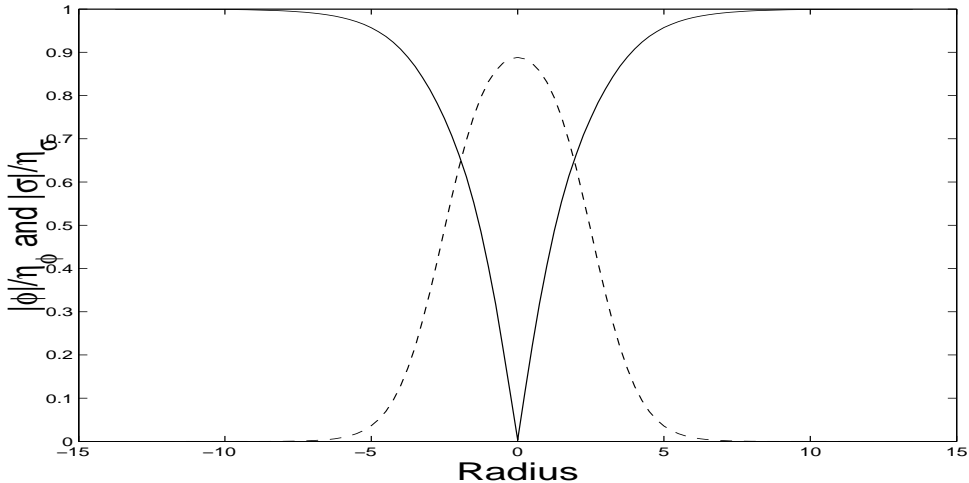}}
\caption{Profiles of $\phi$ and $\sigma$ in the modified Witten model,
plotted in units of
$\eta_{\phi}$ and $\eta_{\sigma}$; the parameters here are
$\lambda_{\phi} = 0.5$, $\eta_{\phi} = 1.0$, $\lambda_{\sigma} = 18.0$,
$\eta_{\sigma} = 0.35$, with the new interaction parameter $\beta' =
3.3$}
\label{phi6}
\end{figure}

To overcome this difficulty, we have modified the interaction term in
eq.(\ref{lagrangian}), $V_{int} = \beta|\phi|^2|\sigma|^2$, by 
considering a toy model with $V_{int} =
\beta'|\phi|^6|\sigma|^2$. Because of the higher power of $\phi$
involved, the effective
potential seen by the condensate is much closer to a square well, which 
allows $\sigma$ to build up a higher and more robust condensate, as can be
seen from fig.\ref{phi6}.  This represents only a quantitative, rather than
qualitative, modification of the vorton model with the new criterion for 
the splitting instability becoming less stringent. Then, the
constraint given by (\ref{condf}) is relaxed to $R / \delta_{\sigma}
\lsim 5$, and we can create numerical configurations in which the
condensate and the underlying vortex are held together tightly.

\section{Vorton simulations}

The theory defined in the previous section is ideal for numerical
simulations. Our code evolves the fields according to the full 3D
equations of motion arising from eq.(\ref{lagrangian}), using a
lattice-inspired hamiltonian formalism \cite{MMR:NR} modified from
\cite{JM:PHD}. energy and charge conservation in all simulations was
maintained to below 1\% accuracy.

First, we study the perfectly
circular chiral case. The initial configuration was obtained using an SOR
relaxation routine to calculate the radial profiles. We then used the ansatz:
\begin{equation}
\sigma(r, \theta, \phi) = |\sigma|(r, \theta)e^{i(k\phi + \omega t)} \, ,
\end{equation}
which describes a homogeneous chiral vorton.  Here, we have neglected small
corrections due to the curvature.

\begin{figure}
\resizebox{13cm}{6.5cm}{\includegraphics{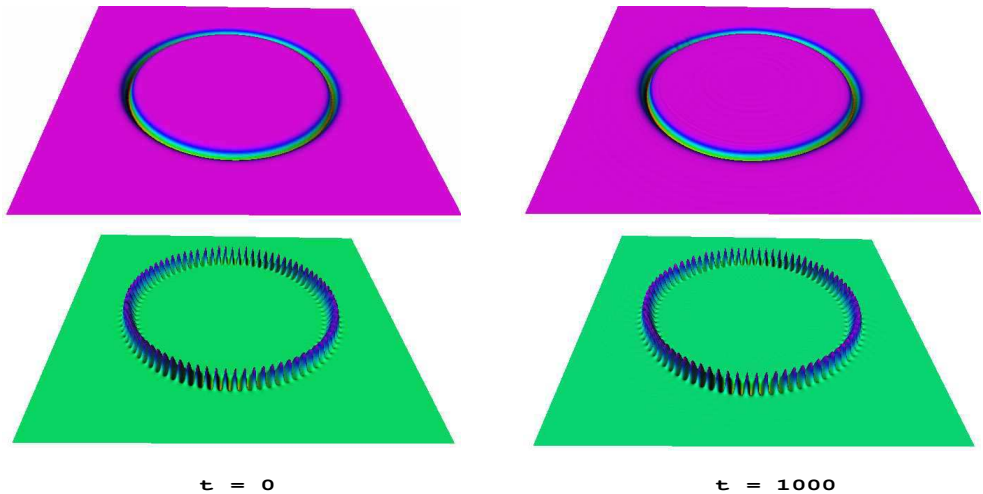}}
\caption{Vorton simulation, showing $|\phi|$ (above) and the real part
of $\sigma$, at $t = 0$ and $t = 1000$ (note the uniform winding of
$\sigma$, meaning a uniform current).}
\label{circle}
\end{figure}

We then let the loop evolve with our code, using (Dirichlet) reflective
boundary conditions which do not act to stabilise the configuration. 
We could observe the loop slowly oscillating around
its equilibrium position, in agreement with the
radial analysis we have given (see also \cite{BPOV:VO,CP:VO,DS:CV,DKPS:CCS}).
As can be seen in fig.\ref{circle}, the
whole structure appears to be remarkably stable.

We now turn our attention to vorton stability with respect to
perturbations in the eccentricity $\epsilon$, by considering a
loop with $\epsilon < 1$. To ensure that the current is initially
homogeneous, we have to consider the modified ansatz:
\begin{equation}
\sigma(r, \theta, s) = |\sigma|(r, \theta)e^{ks + \omega t} \, .
\end{equation}
where $s$ is the arclength along the string. This arclength is given
by an elliptic
integral which we evaluate accurately using the Gauss-Tschebychev algorithm (as
can be found in \emph{e.g.} \cite{NR:NR}). 

\begin{figure}
\includegraphics{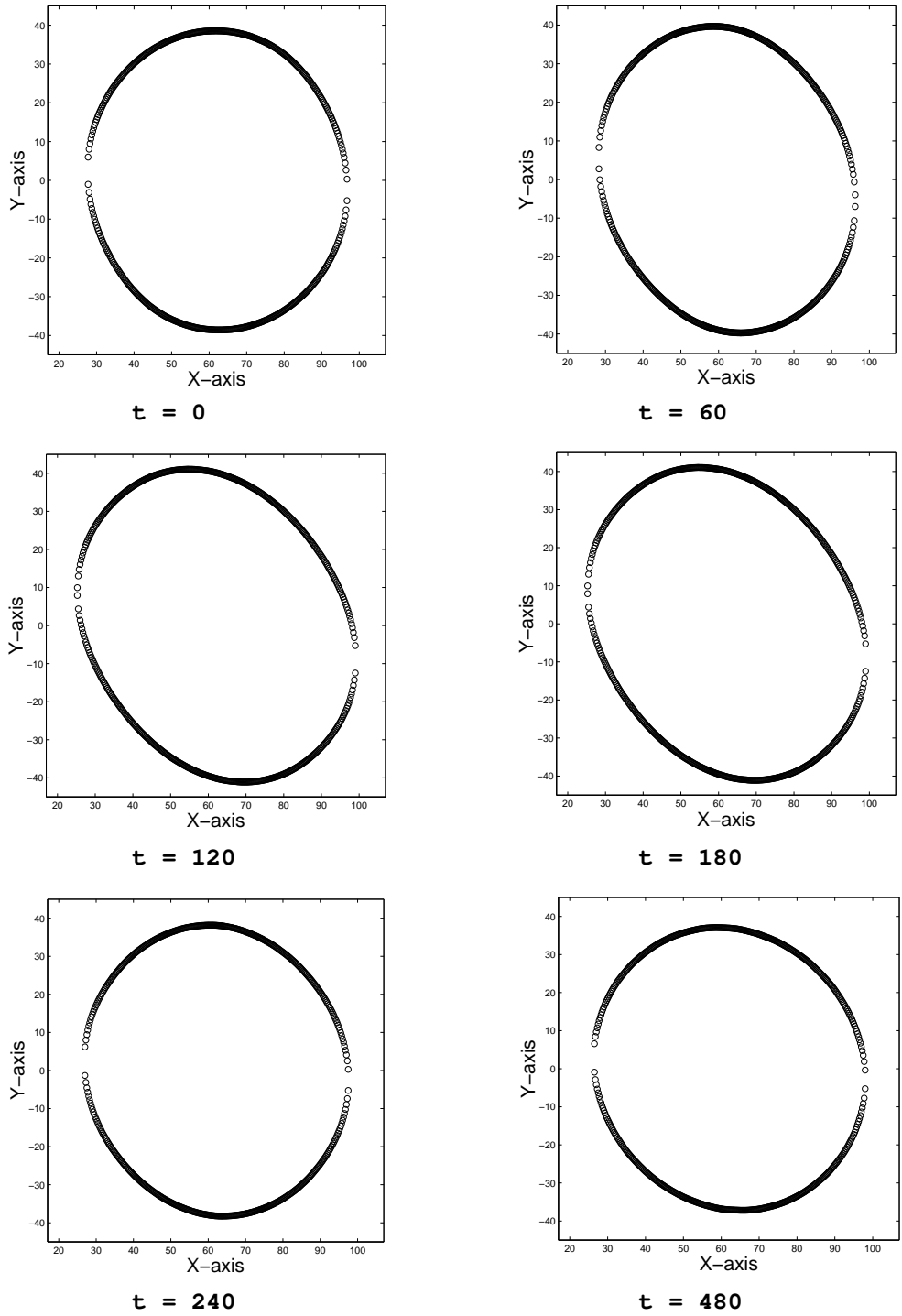}
\caption{Plots of the zeros of $\phi$ in the transverse plane of an
elliptical chiral vorton: the first four plots show the first
period, and the last two are taken after two and three periods
(initially, $\epsilon = 0.4$).}
\label{ellipses}
\end{figure}

Our observations are summarized in fig.~\ref{ellipses}, where we can see clearly
that the loop is actually oscillating between its initial
configuration, and another, somewhat larger loop with higher
eccentricity, tilted in the direction opposite to the current flow. We
note also that this behaviour satisfies $T \simeq L$, where $T$ is the
period of the movement, and $L$ is the length of the loop.

This can be understood if one considers the  effect on the current of
squashing a
circular vorton, to make it look like an ellipse.  To lowest order,
the net effect on 
the angular momentum of the current is to induce
a correction of the form:
\begin{equation}
M = M_0(1+\frac{\epsilon^2}{4}) \, .
\label{M}
\end{equation}
Thus, the angular momentum of the current is incresaed with the eccentricity, 
and this has
to be compensated by a rotation of the loop in the opposite direction
(we discuss
the transfer of momentum at length elsewhere \cite{LS:DCS}).
Despite the oscillating eccentricity, ultimately these loops 
tend to evolve towards more circular
configurations, as can be seen from the last two plots in fig.\ref{ellipses}.

Eccentric loop configurations retained their identity 
for more than 10,000
time steps (many light-crossing times), and so these simulations appear to 
establish that stable vortons
should form during the evolution of the
universe.  Two small caveats to this conclusion remain.  First, in the
very longest 
simulations of over 30,000 timesteps, the build-up of background radiation (due
to the reflective boundary conditions) causes some friction on the
time-varying current
which is eventually driven towards the less stable magnetic regime.
We are developing
absorbing boundary conditions for massive radiation to test the
significance of this boundary artifact.  Secondly, there is the sensitivity of
these objects to their initial conditions, since a slight mismatch in
the phase of the condensate can have dramatic consequences on the subsequent
evolution of the loop. This problem is still under active investigation and 
will be discussed at greater length elsewhere.

\section{Conclusions and discussion}
We have presented in this paper the first simulations of vortons,
which appear to strongly indicate their stability, and hence their cosmological
relevance (as well as in other physical contexts). Stable vortons may
have profound
consequences for cosmology. These objects can be very
massive, and they could  account for dark
matter, or even dominate the universe \cite{MS:VF,CD:CV}. They are naturally associated
with very high energies, and this makes them prime candidates for
high-energy astrophysical puzzles, like cosmic rays \cite{BP:CRV} or
gamma ray bursts \cite{BHV:GRB}. Further study is required to test these
hypotheses, but there can be little doubt that a deeper understanding of the
microphysics of superconducting strings will further the confrontation
between vortons
and observations.

\section*{Acknowledgements}

Y.L. and E.P.S.S. gratefully aknowledge useful conversations with Jose
Blanco-Pillado, Daniele Steer, Brandon Carter and Patrick Peter.
Numerical simulations were performed on the COSMOS supercomputer, the
Origin3800 owned by the UK Computational Cosmology Consortium,
supported by Silicon Graphics Computer Systems, HEFCE and PPARC.

\bibliography{vorton}
\bibliographystyle{unsrt}

\end{document}